\documentclass[prl,twocolumn,superscriptaddress]{revtex4}
\let\onlinecite\cite
\usepackage{graphicx}
\usepackage{amssymb}
\usepackage{amsmath}
\usepackage{amsfonts}
\usepackage{graphicx}
\usepackage{bm}
\newcommand* {\vek}[1]{{\ensuremath{\bm{\mathrm{#1}}}}}
\newcommand* {\kk}{\vek{k}}
\newcommand* {\Ee}{\mathcal{E}}
\newcommand* {\etal}{\textit{et al.}}
\usepackage{array}
\newcolumntype {s}[1]{@{\hspace{#1}}} 
\newcolumntype {L}{>{$}l<{$}}         
\newcolumntype {C}{>{$}c<{$}}         
\newcommand{\marr}[2][@{}l@{}]{\mbox{\renewcommand{\arraystretch}{0.8}%
  \tabcolsep 0pt\begin{tabular}{#1} #2 \end{tabular}}}
\newcommand* {\kdotp}{\ensuremath{\vek{k}\cdot\vek{p}}}

\begin{document}

\title{Robust Level Coincidences in the Subband Structure of Quasi
 2D Systems}

\author{R. Winkler}
\affiliation{Department of Physics, Northern Illinois University,
DeKalb, IL 60115, USA}
\affiliation{Materials Science Division, Argonne National
Laboratory, Argonne, IL 60439, USA}
\affiliation{Department of Electrophysics, National Chiao Tung
University, Hsinchu 30010, Taiwan}

\author{L. Y. Wang}
\affiliation{Department of Electrophysics, National Chiao Tung
University, Hsinchu 30010, Taiwan}

\author{Y. H. Lin}
\affiliation{Department of Electrophysics, National Chiao Tung
University, Hsinchu 30010, Taiwan}

\author{C. S. Chu}
\affiliation{Department of Electrophysics, National Chiao Tung
University, Hsinchu 30010, Taiwan}
\affiliation{National Center for Theoretical Sciences, Physics
Division, Hsinchu 30043, Taiwan}

\date{\today}

\begin{abstract}
  Recently, level crossings in the energy bands of crystals have
  been identified as a key signature for topological phase
  transitions. Using realistic models we show that the parameter
  space controlling the occurrence of level coincidences in energy
  bands has a much richer structure than anticipated previously. In
  particular, we identify robust level coincidences that cannot be
  removed by a small perturbation of the Hamiltonian compatible with
  the crystal symmetry. Different topological phases that are
  insulating in the bulk are then separated by a gapless (metallic)
  phase. We consider HgTe/CdTe quantum wells as a specific example.
\end{abstract}

\pacs{71.20.-b, 73.21.Fg, 03.65.Vf}

\maketitle


Recently level crossings in the energy bands of crystals have become
a subject of significant interest as they represent a key signature
for topological phase transitions induced, e.g., by tuning the
composition of an alloy or the thickness of a quasi two-dimensional
(2D) system \cite{kan05z, mur07, has10, qi10z}. For example, it was
proposed \cite{ber06a} and soon after confirmed experimentally
\cite{koe07, rot09} that HgTe/CdTe quantum wells (QWs) show a phase
transition from spin Hall insulator to a quantum spin Hall regime
when the lowest electron-like and the highest hole-like subbands
cross at a critical QW width of $\sim 65$~{\AA}; see also
\cite{mur07, koe08, liu08, dai08, luo10}. Here we present a
systematic study of level crossings and anticrossings in the subband
structure of quasi 2D systems. We show that the parameter space
characterizing level crossings has a much richer structure than
previously anticipated. In particular, we present examples for
robust level coincidences that are preserved while the system
parameters are varied within a finite range. Similar to the
topological phase transitions characterizing the quantum Hall effect
\cite{tho82}, the insulating $Z_2$ topological phases \cite{kan05z}
thus get separated by a gapless (metallic) phase. Such an additional
phase was previously predicted in Ref.\ \cite{mur07a}. Yet it was
found that this phase could occur only in 3D, but not in 2D. Also,
it was not clear which systems would realize such a phase. Here we
take HgTe/CdTe QWs as a realistic example, though many results are
relevant also for other quasi 2D systems

Level crossings were studied already in the early days of quantum
mechanics \cite{hun27, neu29a, her37z}. They occur, e.g., when atoms
are placed in magnetic fields in the transition region between the
weak-field Zeeman effect and the high-field Paschen-Back effect.
Also, they occur when molecules and solids are formed from isolated
atoms. Hund \cite{hun27} pointed out that adiabatic changes of 1D
sys\-tems---un\-like multi-di\-men\-sion\-al sys\-tems---can\-not
give rise to level crossings. Von Neumann and Wigner \cite{neu29a}
quantified how many parameters need to be varied for a level
crossing. While levels of different symmetries (i.e., levels
transforming according to different irreducible representations,
IRs) may cross when a single parameter is varied, to achieve a level
crossing among two levels of the same symmetry, it is in general
necessary to vary three (two) independent parameters if the
underlying eigenvalue problem is Hermitian (orthogonal).
Subsequently, this problem was revisited by Herring \cite{her37z}
who found that the analysis by von Neumann and Wigner was not easily
transferable to energy bands in a crystal due to the symmetry of the
crystal potential. Similar to energy levels in finite systems,
levels may coincide in periodic crystals if the levels have
different symmetries. Of course, unless the crystal is invariant
under inversion, this can occur only for high-symmetry lines or
planes in the Brillouin zone (BZ), where the group of the wave
vector is different from the trivial group $C_1$. If at one end
point $\kk_1$ of a line of symmetry a band with symmetry $\Gamma_i$
is higher in energy than the band with symmetry $\Gamma_j$, while at
the other end point $\kk_2$ the order of $\Gamma_i$ and $\Gamma_j$
is reversed, these levels cross somewhere in between $\kk_1$ and
$\kk_2$. Herring classified a level crossing as ``vanishingly
improbable'' if it disappeared upon an infinitesimal perturbation of
the crystal potential compatible with all crystal symmetries. In
that sense, a level coincidence at a high-symmetry point of the BZ
such as the $\Gamma$ point $k=0$ becomes vanishingly improbable. For
energy levels with the same symmetry, Herring derived several
theorems characterizing the conditions under which level crossings
may occur. In particular, he found that in the absence of inversion
symmetry level crossings that are \emph{not} vanishingly improbable
may occur for isolated points $\kk$ such that these crossings cannot
be destroyed by an infinitesimal change in the crystal potential,
but they occur at some point near~$\kk$. Here we identify several
examples for such robust level coincidences. This illustrates that
level coincidences in energy bands can be qualitatively different
from level coincidences in other systems \cite{neu29a}.

Recently, several studies focusing on topological phase transitions
recognized the importance of symmetry for level crossings in energy
bands \cite{mur07, koe08, liu08, dai08}. Murakami \etal\
\cite{mur07} studied the phase transition separating spin Hall
insulators from the quantum spin Hall regime, focusing on generic
low-symmetry configurations with and without inversion symmetry.
They found that without inversion symmetry the phase transition is
accompanied by a gap closing at points $\kk$ that are not
high-symmetry points. In inversion symmetric systems the gap closes
only at points $\kk=\vek{G}/2$ where $\vek{G}$ is a reciprocal
lattice vector. Here we show that level crossings in quasi 2D
systems can be characterized by a multitude of scenarios, taking
HgTe/CdTe quantum wells as a specific example for which it is known
that the lowest electron-like and the highest hole-like subbands
(anti)cross for a critical QW width of about $65$~{\AA}
\cite{ber06a, koe07, rot09, pfe00}. In most semiconductors with a
zinc blende structure (point group $T_d$) the $s$-antibonding
orbitals form the conduction band (IR $\Gamma_6$ of $T_d$), whereas
the $p$-bonding orbitals form the valence band ($\Gamma_8$ and
$\Gamma_7$ of $T_d$). The curvature of the $\Gamma_6$ band is thus
positive whereas it is negative for the $\Gamma_8$ band. For finite
$\kk$, the fourfold degenerate $\Gamma_8$ states (effective spin
$j=3/2$) split into so-called heavy hole (HH, $m_z=\pm 3/2$) and
light hole (LH, $m_z=\pm 1/2$) branches. In HgTe, the order of the
$\Gamma_8$ and $\Gamma_6$ bands is reversed: $\Gamma_6$ is located
below $\Gamma_8$ and it has a negative (hole-like) curvature,
whereas $\Gamma_8$ splits into an electron ($m_z=\pm 1/2$) and a
hole ($m_z=\pm 3/2$) branch \cite{dor83z}. HgTe and CdTe can be
combined to form a ternary alloy Hg$_x$Cd$_{1-x}$Te, where the
fundamental gap $E_0$ between the $\Gamma_6$ and $\Gamma_8$ bands
can be tuned continuously from $E_0 = +1.6$~eV in CdTe to $E_0 =
-0.3$~eV in HgTe with a gapless material for $x \approx 0.16$
\cite{dor83z}. Tuning the material composition $x$ thus allows one
to overcome Herring's conclusion \cite{her37z} that a degeneracy at
$k=0$ between two levels of different symmetries is, in general,
vanishingly improbable.

Layers of HgTe and CdTe can also be grown epitaxially on top of each
other to form QWs. At the interface the corresponding states need to
be matched appropriately. The opposite signs of the effective mass
inside and outside the well result in eigenstates localized at the
interfaces \cite{lin85}. We calculate these eigenstates as well as
the corresponding subband dispersion $E_\alpha(\kk)$ using a
realistic $8\times 8$ multiband Hamiltonian $\mathcal{H}$ for the
bulk bands $\Gamma_6$, $\Gamma_8$, and $\Gamma_7$, which fully takes
into account important details of $E_\alpha(\kk)$ such as
anisotropy, nonparabolicity, HH-LH coupling, and spin-orbit coupling
both due to bulk inversion asymmetry (BIA) of the zinc blende
structure of HgTe and CdTe as well as structure inversion asymmetry
(SIA) of the confining potential $V(z)$. For details concerning
$\mathcal{H}$ and its numerical solution see
Refs.~\onlinecite{win03, bia-coeffs}. In the following
$\kk=(k_x,k_y)$ denotes the 2D wave vector.

The symmetry group $\mathcal{G}$ of a QW and thus the allowed level
crossings depend on the crystallographic orientation of the surface
used to grow a QW [a (001) surface being the most common in
experiments]. It also depends on whether we have a system without or
with BIA and/or SIA. The resulting point groups are summarized in
Table~\ref{tab:symgroup}. We show below that these different groups
give rise to a rich parameter space for the occurrence of level
coincidences. For a proper symmetry classification we project the
eigenstates of $\mathcal{H}$ onto the IRs of the respective point
group \cite{bir74}. In the following, all IRs are labeled according
to Koster \etal\ \cite{kos63}. As spin-orbit coupling plays a
crucial role for BIA and SIA \cite{win03} as well as for topological
phase transitions \cite{kan05z, mur07, has10, qi10z}, all IRs
referred to in this work are double-group IRs. For comparison,
Table~\ref{tab:symgroup} also lists the point groups if the
prevalent axial (or spherical) approximation is used for
$\mathcal{H}$. In this approximation, BIA is ignored and different
surface orientations become indistinguishable.

\begin{table}
  \caption{\label{tab:symgroup} The point group of a QW for
   different growth directions starting from a bulk semiconductor with
   diamond structure (point group $O_h$) or zinc blende structure (point
   group $T_d$) for a system without (``sym.'') or with
   (``asym.'') SIA.}
  \renewcommand{\arraystretch}{1.1}
  \begin{tabular*}{\columnwidth}{@{\extracolsep\fill}Cl*{7}{C}} \hline\hline
   & & [001] & [111] & [110] & [mmn] & [0mn] & [lmn] 
   & \marr{axial \\ appr.} \\ \hline
   O_h & sym.  & D_{4h} & D_{3d} & D_{2h} & C_{2h} & C_{2h} & C_{i} & D_{\infty h} \\
       & asym. & C_{4v} & C_{3v} & C_{2v} & C_{s}  & C_{s}  & C_{1} & C_{\infty v} \\
   T_d & sym.  & D_{2d} & C_{3v} & C_{2v} & C_{s}  & C_{2}  & C_{1} & D_{\infty h} \\
       & asym. & C_{2v} & C_{3v} & C_{s}  & C_{s}  & C_{1}  & C_{1} & C_{\infty v} \\
       \hline\hline
   \end{tabular*}
\end{table}

\begin{figure}[tbp]
\centerline{\includegraphics[width=0.99\columnwidth]{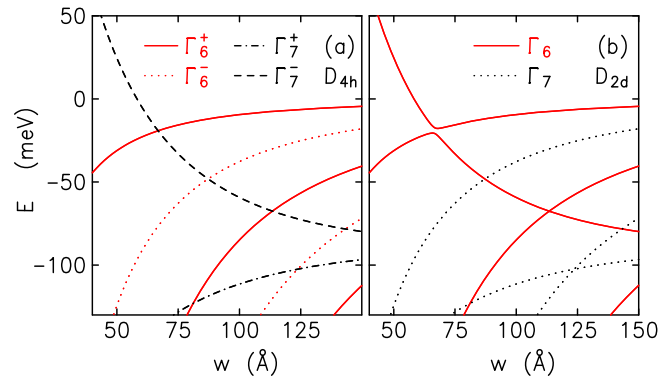}}
\caption{\label{fig:w-scan}(Color online) Subband states in a
 symmetric HgTe/CdTe quantum well (for $k = 0$) as a function of
 well width $w$ calculated with an $8\times 8$ Hamiltonian (a)
 neglecting BIA (point group $D_{4h}$) and (b) with BIA ($D_{2d}$).
 States transforming according to $\Gamma_6^\pm$ of $D_{4h}$
 ($\Gamma_6$ of $D_{2d}$) are shown in red; states shown in black
 transform according to $\Gamma_7^\pm$ of $D_{4h}$ ($\Gamma_7$ of
 $D_{2d}$).}
\end{figure}

First we neglect the small terms in $\mathcal{H}$ due to BIA so that
the bulk Hamiltonian has the point group $O_h$. In the absence of
SIA, a quasi 2D system grown on a (001) surface has the point group
$D_{4h}$ (which includes inversion) and all electron and hole states
throughout the BZ are two-fold degenerate \cite{bir74}. Subband
edges $k = 0$ in a HgTe/CdTe QW as a function of well width $w$ are
shown in Fig.~\ref{fig:w-scan}(a). The HH states transform according
to $\Gamma_6^\pm$ of $D_{4h}$. The electron-like and LH-like
subbands transform according to $\Gamma_7^\pm$. As expected, the
$\Gamma_6^\pm$ and $\Gamma_7^\pm$ subbands may cross as a function
of $w$.

In the presence of SIA we cannot classify the eigenstates anymore
according to their behavior under parity. Without BIA the point
group becomes $C_{4v}$. HH states transform according to $\Gamma_6$
of $C_{4v}$ and electron- and LH-like states transform according to
$\Gamma_7$. The level crossings depicted in Fig.~\ref{fig:w-scan}(a)
remain allowed in this case \cite{koe08, rot10}.

The situation changes when taking into account BIA. Without SIA the
point group becomes $D_{2d}$. In this case, all subbands transform
alternately according to the IRs $\Gamma_6$ and $\Gamma_7$ of
$D_{2d}$, irrespective of the dominant spinor components. In
particular, both the highest HH state and the lowest conduction band
state transform according to $\Gamma_6$ of $D_{2d}$ so that around
$w \simeq 65$~{\AA} we obtain an anticrossing between these levels
of about $2.9$~meV (for $k=0$), see Fig.~\ref{fig:w-scan}(b)
\cite{koe08, liu08, dai08}.
With both BIA and SIA the point group becomes $C_{2v}$. Now we have
only one double group IR $\Gamma_5$. Thus it follows readily that
all subbands anticross as a function of a continuous parameter such
as the well width.

\begin{figure}[tbp]
\centerline{\includegraphics[width=0.99\columnwidth]{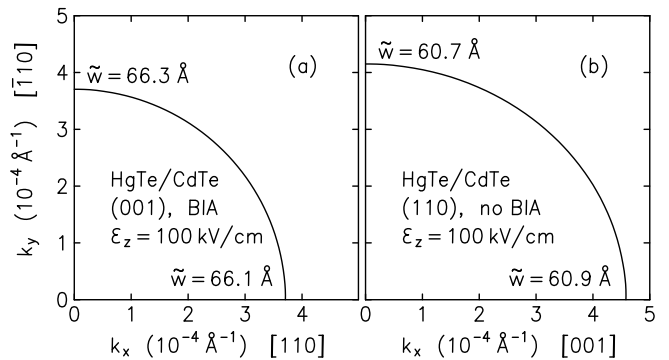}}
\caption{\label{fig:kloop}Critical wave vectors $\tilde{\kk}$ that
 give rise to a level coincidence in a HgTe/CdTe QW (a) on a (001)
 surface taking into account BIA (b) on a (110) surface neglecting
 BIA. In both cases a perpendicular field $\Ee_z = 100$~kV/cm was
 assumed. In (a) the level coincidence requires a well width
 $\tilde{w} = 66.1$~{\AA} for $\tilde{\kk} \parallel [110]$ and
 $\tilde{w} = 66.3$~{\AA} for $\tilde{\kk} \parallel [\bar{1}10]$.
 In (b) we have $\tilde{w} = 60.9$~{\AA} for $\tilde{\kk} \parallel
 [001]$ and $\tilde{w} = 60.7$~{\AA} for $\tilde{\kk} \parallel
 [\bar{1}10]$.}
\end{figure}

While BIA opens a gap at $k = 0$, level coincidences remain possible
for some $\tilde{\kk} \ne 0$ when the well width $w$ is tuned to a
critical value $\tilde{w}$ \cite{mur07, her37z}. Considering a (001)
surface with BIA, we find, indeed, that for each direction $\phi$ of
$\kk = (k, \phi)$, critical values $\tilde{w}$ and $\tilde{k}$ exist
that give rise to a band crossing. Thus we get a line in $\kk$ space
where the bands cross when $w$ is varied within some finite range.
This result holds for QWs on a (001) surface with BIA, without and
with SIA (as studied experimentally in Refs.\ \cite{koe07, rot09}).
As an example, Fig.~\ref{fig:kloop}(a) shows $\tilde{\kk}$ in the
presence of a perpendicular electric field $\Ee_z = 100$~kV/cm.

In general, three independent parameters must be tuned for a level
coincidence in a quantum mechanical systems \cite{neu29a} if the
underlying eigenvalue problem is Hermitian. While the multiband
Hamiltonian $\mathcal{H}$ used here \cite{win03} is likewise
Hermitian (not orthogonal), only two independent parameters ($w$ and
$k = |\kk|$) are necessary to achieve the level degeneracy. We have
here an example for the robustness of band coincidences under
perturbations that was predicted by Herring \cite{her37z} to occur
in systems without a center of inversion (in multiples of four). It
shows that level coincidences in energy bands can behave
qualitatively different from level coincidences in other quantum
mechanical systems \cite{neu29a}. We note that the band coincidences
found here are not protected by symmetry in the sense that---unlike
the other cases discussed above---the group of $\tilde{\kk}$ is the
trivial group $C_1$ containing only the identity.

The situation is different for quasi 2D systems grown on a (111)
surface. In the absence of BIA and SIA, the point group is $D_{3d}$.
HH states at $k=0$ transform according to the complex conjugate IRs
$\Gamma_5^+ \oplus \Gamma_6^+$ or $\Gamma_5^- \oplus \Gamma_6^-$,
where $\oplus$ indicates that these IRs must be combined due to time
reversal symmetry. All other subband edges transform according to
$\Gamma_4^\pm$. In the presence of BIA and/or SIA the point group
becomes $C_{3v}$. Then HH states transform according to the complex
conjugate IRs $\Gamma_5 \oplus \Gamma_6$. Electron-like and LH-like
states transform according to $\Gamma_4$. Thus it follows that on a
(111) surface the HH states always cross the other states at $k=0$
as a function of $w$ [similar to Fig.~\ref{fig:w-scan}(a)].
The IRs for different geometries starting out from a (001) or (111)
surface are summarized in Table~\ref{tab:ireps}.


\begin{table}[tbp]
  \caption{\label{tab:ireps} Irreducible representations of quasi 2D
   states ($\kk=0$) on a (001) and (111) surface, starting from a bulk
   semiconductor with diamond (point group $O_h$) or zinc blende
   (point group $T_d$) structure for a system without (``sym.'') or with
   (``asym.'') structure inversion asymmetry.}
  \centering
  \tabcolsep 1.7ex
  \renewcommand{\arraystretch}{1.2}
  \begin{tabular}{@{}Ll*{2}{s{2\tabcolsep}*{3}{C}}@{}}
    \hline \hline
    & & \multicolumn{3}{c}{(001)} & \multicolumn{3}{c@{}}{(111)} \\
    \multicolumn{1}{@{}l}{\makebox[0pt][l]{bulk}} &
    & \makebox[0pt]{group} & \makebox[0pt]{c, LH} & \makebox[0pt]{HH} 
    & \makebox[0pt]{group} & \makebox[0pt]{c, LH} & \makebox[0pt]{HH}
    \\ \hline
    O_h & sym. & D_{4h} & \Gamma_7^\pm & \Gamma_6^\pm
    & D_{3d} & \Gamma_4^\pm & \Gamma_5^\pm \oplus \Gamma_6^\pm \\
        & asym. & C_{4v} & \Gamma_7 & \Gamma_6
    & C_{3v} & \Gamma_4 & \Gamma_5 \oplus \Gamma_6 \\
    T_d & sym. & D_{2d} & \Gamma_{7/6} & \Gamma_{6/7}
    & C_{3v} & \Gamma_4 & \Gamma_5 \oplus \Gamma_6 \\
        & asym. & C_{2v} & \Gamma_5 & \Gamma_5
    & C_{3v} & \Gamma_4 & \Gamma_5 \oplus \Gamma_6 \\
    \hline \hline
  \end{tabular}
\end{table}


Finally we consider quasi 2D states on a (110) surface. In the
absence of BIA and SIA, the point group becomes $D_{2h}$. Here, all
subbands transform alternately according to $\Gamma_5^+$ and
$\Gamma_5^-$ with the topmost HH-like subband being $\Gamma_5^+$ and
the lowest electron-like subband being $\Gamma_5^-$. A level
crossing as a function of $w$ is thus again allowed at $k = 0$. In
the presence of either BIA or SIA the symmetry is reduced to
$C_{2v}$. While the point group in both cases is the same
\cite{110-c2v}, we obtain a remarkable difference between these
cases. With SIA the level crossing occurs for a line in $\kk$ space,
similar to the (001) surface, see Fig.~\ref{fig:kloop}(b). With BIA
we obtain a level crossing only for $\kk \parallel [\overline{1}10]$
with $\tilde{k} \approx 0.0012$~\AA$^{-1}$ and $\tilde{w} \approx
62.5$~{\AA}, thus giving an example for the level crossings
occurring for isolated points $\tilde{\kk} \ne 0$ as discussed by
Murakami \etal\ \cite{mur07}. These examples illustrate that the
occurrence of level crossings at either isolated points or along
continuous lines in parameter space is not simply related with the
system symmetry \cite{110-c2v}. In the presence of both BIA and SIA
(group $C_s$) we have the same situation as with BIA only, i.e.,
adding SIA changes the values of $\tilde{k}$ and $\tilde{w}$, but we
keep $\tilde{\kk} \parallel [\overline{1}10]$.


In conclusion, we have shown that a rich parameter space
characterizes the occurrence of level coincidences in the subband
structure of quasi 2D systems. In particular, we have identified
level coincidences for wave vectors $\tilde{\kk} \ne 0$ that cannot
be removed by a small perturbation of the Hamiltonian compatible
with the QW symmetry \cite{her37z}. Taking into account the full
crystal symmetry of real materials is an important difference
between the current analysis and previous work that considered only
lattice periodicity, inversion and time reversal symmetry. The full
set of symmetries imposes additional constraints on the band
Hamiltonian beyond the torus topology of the BZ that reflects the
translational symmetry. These additional constraints generally
reduce the number of parameters that are required to obtain level
crossings \cite{her37z} so that robust level coincidences can be
achieved even in quasi 2D systems. As quasi 2D systems can be
designed and manipulated in various ways not available in 3D this
opens new avenues for both experimental and theoretical research of
topologically nontrivial materials.

As a specific example, we have considered HgTe/CdTe QWs, where a
particular level crossing reflects a topological phase transition
from spin Hall insulator to a quantum spin Hall regime \cite{ber06a,
 koe07, rot09}. The robustness of the level coincidences found here
implies that these phases, which are insulating in the bulk, are
separated by a gapless phase similar to the metallic phases that
separate the insulating quantum Hall phases \cite{tho82}. While in
HgTe/CdTe QWs the range of critical well widths $\tilde{w}$ giving
rise to the metallic phase is rather small (about 0.1 monolayers),
we expect that future research will be able to identify materials
showing larger parameter ranges that can be probed more easily in
experiments. We note that our symmetry-based classification of level
crossings is independent of specific numerical values of the band
structure parameters entering the Hamiltonian $\mathcal{H}$. Indeed,
our findings are directly applicable also to other quasi 2D systems
made of bulk semiconductors with a zinc blende or diamond structure
such as hole subbands in GaAs/AlGaAs and SiGe quantum wells. In
general, the $\kdotp$ coupling between the LH1 ($\Gamma_7^+$ of
$D_{4h}$) and HH2 ($\Gamma_6^-$) subbands gives rise to an
electron-like dispersion of the LH1 subband for small wave vectors
$k$ \cite{bro85}. If these subbands become (nearly) degenerate at
$k=0$, the coupling between these subbands becomes the dominant
effect. This situation is described by the same effective
Hamiltonian that characterizes the subspace consisting of the lowest
electron and highest HH subband in a HgTe/CdTe QW \cite{ber06a}. It
can be exploited if biaxial strain is used to tune the separation
between the LH1 and HH2 subbands \cite{voi84}.

\begin{acknowledgments}
  RW appreciates stimulating discussions with T.~Hirahara,
  A.~Hoffmann, L.~W. Molenkamp, and S.~Murakami. He thanks the Kavli
  Institute for Theoretical Physics China at the Chinese Academy of
  Sciences for hospitality and support during the early stage of
  this work. This work was supported by Taiwan NSC (Contract No.\
  99-2112-M-009-006) and a MOE-ATU grant. Work at Argonne was
  supported by DOE BES under Contract No.\ DE-AC02-06CH11357.
\end{acknowledgments}


\begin{thebibliography}{10}
\bibitem{110-c2v}
2D systems on a (110) surface with either BIA or SIA have the point group
  $C_{2v}$. Yet with SIA the symmetry axis of $C_{2v}$ is perpendicular to the
  (110) plane, whereas with BIA this axis is along the in-plane $[001]$ axis.
\end{thebibliography}
\end{document}